# Narrow-gap Semiconducting Superhard Amorphous Carbon with Superior Toughness

Shuangshuang Zhang[1,8], Yingju Wu[1,8], Kun Luo[1,2,8], Bing Liu[1,8], Yu Shu[1], Yang Zhang[1,2], Lei Sun[1], Yufei Gao[1,2], Mengdong Ma[1], Zihe Li[1], Baozhong Li[1], Pan Ying[1,2], Zhisheng Zhao[1,9,*], Wentao Hu[1], Vicente Benavides[3,4], Olga P. Chernogorova[5], Alexander V. Soldatov[1,6,7,*], Julong He[1], Dongli Yu[1], Bo Xu[1], Yongjun Tian[1,*]

[1]Center for High Pressure Science (CHiPS), State Key Laboratory of Metastable Materials Science and Technology, Yanshan University, Qinhuangdao, Hebei 066004, China

[2]Hebei Key Laboratory of Microstructural Material Physics, School of Science, Yanshan University, Qinhuangdao 066004, China

[3]Department of Engineering Sciences and Mathematics, Luleå University of Technology, SE-97187 Luleå, Sweden

[4]Department of Materials Science, Saarland University, Campus D3.3, D-66123, Saarbrücken, Germany

[5]Baikov Institute of Metallurgy and Materials Science, Moscow 119334, Russia

[6]Center for High Pressure Science and Technology Advanced Research, Shanghai 201203, China

[7]Department of Physics, Harvard University, Cambridge, MA 02136, USA

[8]These authors contributed equally to this work

[9]Lead Contact

*Corresponding authors: zzhao@ysu.edu.cn (Z.Z.), Alexander.Soldatov@ltu.se (A.S.) or fhcl@ysu.edu.cn (Y.T.)

**SUMMARY**

New carbon forms exhibiting extraordinary physico-chemical properties can be generated from nanostructured precursors under extreme pressure. Nevertheless, synthesis of such fascinating materials is often not well understood that results, as is the case of $C_{60}$ precursor, in irreproducibility of the results and impeding further progress in the materials design. Here the semiconducting amorphous carbon having bandgaps of 0.1-0.3 eV and the advantages of isotropic superhardness and superior toughness over single-crystal diamond and inorganic glasses are produced from transformation of fullerene at high pressure and moderate temperatures. A systematic investigation of the structure and bonding evolution was carried out by using rich arsenal of complimentary characterization methods, which helps to build a model of the transformation that can be used in further high p,T synthesis of novel nanocarbon systems for advanced applications. The produced amorphous carbon materials have the potential of demanding optoelectronic applications that diamond and graphene cannot achieve.

**INTRODUCTION**

Given structural diversity of the allotrope forms (graphite, graphene, diamond, fullerene and nanotubes, nano-onions, etc.), carbon exhibits it is likely the most intriguing element of the periodic table. Bonding variety - $sp$, $sp^2$, $sp^3$ hybridization - is manifested in distinct electronic and mechanical properties. For example, diamond has a three-dimensional structure comprised of rigid $sp^3$ bonds, while graphite is a stack of weakly bonded (by van der Waals forces) graphene layers with honeycomb lattice of covalently bonded $sp^2$ carbons. As a result, diamond is a superhard insulator, while graphite is a soft semi-metal. A free-standing single-layer graphene is mechanically flexible with ultra-high carrier mobility. However, the gapless character of graphene makes it unsuitable for an on/off operation in field-effect transistors, thus hindering it from becoming the "base" for the next-generation electronic devices.[1] Considerable efforts have been made to open a sizable bandgap in graphene via doping, adatoms, quantum confinement on nanometer scale as well as symmetry breaking by an external electric field or applying high pressure.[2] So far, the bandgap in bilayer graphene achieved at ambient conditions, can reach several hundreds of meV thus bringing a new hope on its application in electronic devices.[3]

The carbon forms in two or more hybrid states show unusual electronic and mechanical properties combining the advantages of each state, so that they have always been sought.[4] Most typical examples are the $sp^2$-$sp^3$ amorphous carbon films, i.e. a-C(:H) and ta-C(:H), with thickness of a few hundred nanometers produced by various deposition techniques.[4–8] The amorphous carbon films are semiconductors with a bandgap in the range of ~0.4-3 eV, possess outstanding mechanico-tribological properties (superhigh hardness and excellent wear resistance),[4,6,7] and, therefore, widely used as superior protective coatings and optoelectronic applications.[4] The semiconducting properties of amorphous carbon films are precious, but reducing the bandgap of the amorphous carbon films below 0.4 eV is a very challenging task, as it's accompanied by a significant reduction of the mechanical properties.[8] Besides, high internal stresses that inevitable accompany the films manufacturing often lead to poor adhesion to the substrate and weaken durability and life time of the films and also prevent synthesis of these materials in a bulk form.[5] Nevertheless, a solution to this major problem may be found by using the other manufacturing protocol that employs, for example, high pressure technique. Indeed, much hope and enthusiasm was brought about by scientific reports highlighting extraordinary properties of the disordered carbon systems derived from various $sp^2$ carbon precursors, typically including fullerene[9–13] and glassy carbon [14,15], under high pressure. For example, a class of lightweight amorphous carbon bulk materials displaying the robust elastic recovery after indentation, electro-conductive, high strength and hardness were recently obtained by compressing glassy carbon at pressures of 5-25 GPa and high temperatures ≤ 1200 ℃.[14,16] These materials are composed of disordered multi-layer graphene sheets, which are locally buckled or linked by $sp^3$ carbon nodes, resulting in the remarkable combination of the mechanical properties they exhibit. Another class of superelastic hard amorphous carbon material that is built of disordered nano-sized graphene clusters can be produced by crushing $C_{60}$ molecules at pressures of 5-8 GPa and temperatures > 800 ℃.[11,17,18] It was proposed that the correlation between the graphene nanoclusters orientation and crystal structure of the parent fullerene endows the unique mechanical properties.[11] Similarly to a-C(:H) and ta-C(:H) coatings, these materials provide tremendous enhancement of the tribological properties as additives in composites - 10-fold and 2-fold decrease of wear resistance and the friction coefficient, respectively.[19]

Increase of the synthesis pressure to 12-15 GPa brought up more exciting results on the $C_{60}$ transformation, and some reports claimed the synthesized amorphous carbon materials with ultrahigh hardness can scratch single-crystal diamond.[12,20] Regretfully, the reports on mechanical properties of these amorphous carbon materials were often controversial and/or inconclusive.[10,18,21] Besides, there were only occasional and insufficient studies of their electrical and optical properties,[22] and no direct characterization of the structural transformations on nanoscale and their dependence in the synthesis pressure and temperature (p,T) has been conducted to date. Differences in the synthesis p,T and process protocol (the rate of temperature and pressure rise/drop, the holding time at target p,T) as well as the pressure hydrostaticity level in the previous experiments strongly complicate comparison and ascertaining the data in a systematic way and may also be the reasons behind poor reproducibility of the produced samples.[10] Consequently, even general understanding of these amorphous carbon materials synthesis is less than complete. There was only one valuable attempt to model the behavior of $C_{60}$ precursor into the amorphous carbon materials produced at high p,T - Brazhkin et al.[22] proposed mechanism of 3-dimensional (3D) polymerization of $C_{60}$ as a two stage process, i.e. the initial polymerization via 2+2 cycloaddition mechanism followed by "heavy" level of polymerization with four or more covalent bonds ($sp^3$ atoms) bridging neighboring buckyballs. Nevertheless, that effort was limited to the formation of 3D-$C_{60}$ polymers and, to the best of our knowledge, no structural model of the fullerene transformation to various amorphous carbon materials describing their structure dependence on the synthesis parameters has been proposed to date. That strongly impedes the materials design in the field and prospective of their practical application. We have therefore undertaken the present study to fill this gap by bringing to use the broad array of complimentary analytical tools to characterize a wide range of physical properties of the amorphous carbon materials recovered after the high p,T synthesis.

Here we present the results of a systematic study of $C_{60}$ fullerene transformation to amorphous carbon materials at 15 GPa and temperatures exceeding the threshold of the buckyballs integrity. The synthesis pressure was selected slightly above the pressure level (12-13 GPa) at which the most interesting[10,12,21] and controversial[17] results were obtained before. A broad array of complimentary analytical tools including X-ray diffraction (XRD), Raman spectroscopy, high-resolution transmission electron microscopy (HRTEM), and electron energy-loss spectroscopy (EELS) were used to reveal the key stages of the structure evolution. Especially with the help of HRTEM observation, the subtle differences in structure can be identified. Mechanical properties of the amorphous carbons have been studied by using three independent methods, i.e., Vickers, Knoop and nano-indentation scales to assure reliability and full compatibility of the results and revealed diamond-like hardness of the synthesized materials and fracture toughness comparable to that of nanodiamond.[23] We found that the produced amorphous carbon materials are the semiconductors with bandgaps of only ~0.1-0.3 eV that were derived from Fourier transform infrared-attenuated total reflectance (ATR-FTIR)) spectra and temperature-dependent electrical resistivity. This kind of narrow-gap amorphous carbon semiconductors with superior mechanical properties may have unique photoelectric applications, such as in the mid-far infrared field.

**RESULTS AND DISCUSSION**

**Characterization of Microstructure**

The XRD patterns in Figure 1A reveal the phase transition of $C_{60}$ after treatment of high pressure of 15 GPa and different temperatures ranging from 550 to 1200 ℃. One can see that the raw $C_{60}$ fullerene first transforms into 3D-$C_{60}$ accompanied by a certain degree of amorphization at the relative low synthesis temperatures T = 550 ℃ (sample ①) following the similar rules previously reported,[12,24] and the amorphization is almost finished at T = 700 ℃ (sample ②). The amorphous carbon materials ③ and ④ (depicted as *am*-I and *am*-II) were quenched from 800 and 1000 ℃, respectively, and they possess three main broad diffraction peaks around structure factor q = ~2.1, ~3.0 and ~5.4 Å$^{-1}$ with close positions but different intensity, especially for the first two peaks. The peak at ~2.1 Å$^{-1}$ is corresponding to the average interlayer spacing (*d* = ~2.99 Å) of the compressed disordered graphite-like component in amorphous carbon, similar to that of *sp²-sp³* compressed glassy carbon[14,25] and so-called compressed graphite formed after nonhydrostatic pressurization of $C_{60}$ at ambient temperature.[26] With the synthesis temperature increase from 700 to 1000 ℃, the diffraction peak at ~2.1 Å$^{-1}$ is gradually narrowed and slightly shifted to the right position, meaning that the layers in the compressed disordered graphite-like component are more regular and the average interlayer spacing becomes smaller. In addition, it is found that the ~2.1 Å$^{-1}$ peak intensity increases first and then decreases. This abnormal change should be related to the complete fragmentation of 3D-$C_{60}$, re-polymerization of small graphene fragments with different sizes and shapes, and the further development of graphite-like and diamond-like disordered components in microstructure, which are supported by the following HRTEM observation on the microstructure change during the phase transformation. When the synthesis temperature is increased to 1200 ℃, the recovered samples ⑤ show four sharp diffraction peaks, implying crystallization. Among them, the first peak at q = ~2.0 Å$^{-1}$ corresponds to the previously so-called compressed graphite,[25,26] and the last three peaks at ~3.02, ~4.99, and ~5.84 Å$^{-1}$ are corresponding to the diamond reflections of (111), (220) and (311), respectively.

Figures 1B, 1C and Figures S1-S3 show the Raman spectra of recovered samples. Position of the Raman modes in the spectrum of the raw $C_{60}$ (Figure S1) are in good agreement with the literature data, and the strongest peak at 1469 cm$^{-1}$ is corresponding to the $A_{2g}$ pentagon-pinch mode in monomeric fullerene.[27] After high-pressure (HP) and high-temperature (HT) treatment, the Raman spectra exhibit profound change in the recovered materials. The samples ①-③ reveal a very broad asymmetric feature between 1200-1800 cm$^{-1}$ with highest intensity and between 1500-1600 cm$^{-1}$ that is typical for highly disordered or amorphous carbon systems and manifests the G-band, characteristic of vibrational modes of $sp^2$-bonded carbon atoms. Raman spectrum of the sample ④ exhibits a shoulder developing on the low frequency side in the broad main feature of the spectrum. This shoulder is growing in intensity with further synthesis temperature increase in sample ④ and reaches the intensity of the main feature at ~1600 cm$^{-1}$. A tremendous change in the Raman spectrum profile is demonstrated by the spectrum of the sample ⑤ synthesized at 1200 ℃, the highest temperature of the experiment - two separate peaks that are narrower than the correspondent features in the spectrum ④ we assign to D and G vibrational bands of the honeycomb structure of graphene layers. Specific information about the disordered carbon structures can be obtained from the peak decomposition of the Raman spectra and their assignment and analysis. Table S1 summarizes the peaks assignment used in our model: G-band[28–31] ($E_{2g}$ Raman mode) characteristic of various structural forms of $sp^2$-bonded carbon atoms; vibrational modes of 5-fold[4–6,11,12] and 7-fold[11,32,33] aromatic rings relevant to our systems given the precursor material, $C_{60}$; D-band[11,28–31,34] describing the "breathing" vibrational mode of 6-fold (hexagonal) aromatic "rings" that is also $A_{1g}$ Raman mode in graphene associated with defects as is the other double resonant, defects-related graphene mode D';[11,28–31] and finally, the characteristic Raman features of $sp^3$-bonded carbons - T-band,[29] the hallmark of disordered, nearly pure $sp^3$-bonded carbon systems (ta-C(:H)) and nanodiamond and a peak that has been reported for amorphous $sp^2$-$sp^3$ carbon systems which we term here as "disorder-$sp^3$" (d-$sp^3$) peak.[35,36]

A Breit-Wigner-Fano (BWF) function, the asymmetric ("stretched") Lorentzian, is suitable for description of the G-band, the contribution from $sp^2$-bonded carbon atoms in amorphous and disordered carbon systems,[28–31] because of complexity and structural inhomogeneity of the $sp^2$ component that may include various structural units present simultaneously in the system - linear chains, fused aromatic rings, and, as a specific case of the latter, small clusters/stacks of fused hexagonal rings – a few layer graphene "seeds". Thus we used the BWF function in peak decomposition of highly disordered carbon systems studied in this work. Deviation of the BWF from the Lorentzian peak shape, i.e. the BWF asymmetry degree, is mathematically described by so caller $Q_{BWF}$ factor: the smaller $1/Q_{BWF}$ the less asymmetry BWF peak exhibits with an infinite $Q_{BWF}$ factor (zero $1/Q_{BWF}$) corresponding to Lorentzian peak shape as asymptotic of BWF function. Gaussian functions were used for fitting the F, D, d-$sp^3$ and 7-ring-derived Raman peaks.

The peak decomposition of the Raman spectra is displayed on Figure 1B and more details are shown in tables accompanying Figures S2 and S3 in supplemental information, respectively. The summaries of the fitting results are presented in Table 1 below. For such spectra the $Q_{BWF}$ factor can be directly associated with the $sp^3$ bonded carbon atoms content in the system. Since Raman spectra in our case are composed of a number of peaks, the obtained $Q_{BWF}$ values cannot be directly associated with $sp^3$ carbons content. Nevertheless, from qualitative comparison of the results we can draw a conclusion about higher/lower $sp^3$ carbons fraction in different samples that will result in lower/higher $Q_{BWF}$ values, respectively. In the series of Raman spectra in Figure S2, the shape of BWF peak gradually changes with the synthesis temperature. For example, the $Q_{BWF}$ factor derived from the BWF peak decreases with increasing the synthesis temperature from 550 ℃ in sample ① (Figure S2A) through 1200 ℃ in sample ⑤ (Figure S3B) that implies increase of the $sp^3$ carbons fraction. D-band is characteristic of hexagonal rings in the system. Visually unnoticeable in the spectra collected from the sample ① and ② synthesized at "low" temperatures (550 ℃ and 700 ℃, respectively), the D-band intensity gradually increases with the synthesis temperature that is first manifested in mentioned above appearance of a shoulder at low frequency side of G-band (BWF peak) and, finally, grows into a well-defined separate peak with higher intensity than G-band implying multi-layer graphene (MLG) presence in the system. The $Q_{BWF}$ factors and $1/Q_{BWF}$ parameters derived from the Raman spectra reflect dependence on the samples synthesis temperature (Table 1). Interestingly, the $1/Q_{BWF}$ parameter value (-0.08) of the BWF/G-band peak, in sample ⑤ (1200 ℃) is similar to that reported for G-band in graphene.[37] The BWF peak shape with such $1/Q_{BWF}$ parameter is approaching the symmetrical function (Lorentzian peak shape).

Another important conclusion about the systems structure evolution regarding the $sp^2$ carbons component can be drawn from the D/G peak area ratio that corresponds to the fraction of $sp^2$ carbons incorporated in the hexagonal rings to the total number of the $sp^2$ carbons in a system. Indeed, D/G peak area ratio increases from 0.1 in the *am*-I (800 ℃) through 0.38 in the *am*-II (1000 ℃) to reach finally a value of 1.89 in nanographene/nanographite (1200 ℃). Such evolution of the Raman spectra reflects gradual growth of hexagonal rings in the system at the expense of other $sp^2$ carbon-based structural units until all $sp^2$-bonded carbon atoms are associated with hexagons in a pure hexagonal rings-based (honeycomb) structure - the nanographene/nanographite phase that concludes the structure evolution. Top panel in Figure S4 illustrates this process via dependence of the G-band peak width (G-FWHM) and the D/G peak area ratio on the synthesis temperature. Even though G-FWHM decreases as a function of temperature it remains very broad (>200 cm$^{-1}$) until the temperature reaches 1200 ℃. Such broad G-band corresponds to coherent

Raman scatterers' (fused hexagons/nanographene clusters') size less than ~2 nm. The samples ①-④ (550-1000 ℃) can thus be placed in the stage II of Ferrari and Robertson ternary phase diagram/graphene amorphization/growth pathway.[28,29] Then, as the synthesis temperature increases from 700 to 1000 ℃, the system moves from Stage II towards the Stage I (Figure S4, bottom panel). Another evidence of hexagonal rings dominance among the nanostructural units built of $sp^2$ atoms at certain stage of system evolution comes from comparison of the Raman spectra collected with different laser excitations. Indeed, Figure 1C displays clear blue shift of the BWF (G-band) peak in the spectrum acquired with 325 nm (UV) laser excitation with respect to the one collected with 532 nm (visible) laser in the samples ② and ③ (G-band dispersion) whereas in the sample ④ no G-band dispersion present (the BWF peaks coincide). The G-band dispersion originates in a variety of $sp^2$ carbon-based structural units (chains, different fused aromatic rings, small clusters, etc.) present in an amorphous carbon system that results in a resonant behavior of the G-band as different structural units will be in resonance with different laser excitations thus causing the G-band peak appear at different frequencies. Consequently, the G-band dispersion testifies for high structural inhomogeneity in the system and no preference/dominance of certain structural species, i.e. in the samples ② and ③ there are many structural units (short chains, pentagons, hexagons, heptagons, etc.) created upon fullerene cages collapse present whereas the hexagons do not dominate. On the contrary, lack of dispersion in sample ④ indicates on domination hexagon-based structural elements in the system. In addition, $sp^2$ carbons have low Raman cross-section for UV laser excitation, that is corroborated with low intensity of the UV Raman spectra in the energy range of D-band Figure 1C, and, instead gradual rise of the D-band in the Raman spectra acquired with 532 nm laser must be interpreted as another evidence for hexagons number increase/graphene growth in the system on temperature increase.

Finally, the Raman spectra at 1200 ℃ are associated with the Stage I graphene amorphization/growth trajectory both from D/G (BWF) peak area ratio, G-band peak position, width and peak shape ($Q_{BWF}$ factor value), see Table 1. This is related to establishing nanoclustered MLG/nanographite phase in the system. Nanographene clusters' size ($L_a$) was estimated from the G-band FWHM[38] and D/G peak area ratio ($A_D/A_G$)[39] at 10±2 nm and 9±2 nm, respectively.

HRTEM was used to directly observe the microstructure difference in quenched carbon samples (Figure 2 and Figure S5). Figure 2A shows the intermittent periodic patterns in sample ①, implying the local amorphization caused by the partial collapse of 3D-$C_{60}$ structure, and the corresponding SAED (Figure S5A) and fast Fourier Transform (FFT) patterns show the broadened amorphous halo and crystalline spots together. For *am*-I obtained at higher synthesis temperature, a completely disordered structure can be seen, and inside two kinds of short-range microstructures can be distinguished, one is disordered fingerprint-like small curved fragments with graphite-like interlayer spacing (blue box area in Figure 2B), and the other looks more compact with shorter spacing, which may be due to $sp^3$ linkage in this area (red box area in Figure 2B). The corresponding SAED (Figure S5B) and FFT patterns only show the feature of amorphous halo. With the further increase of synthesis temperature, the previously fingerprint-like small curved fragments in *am*-I are combined into more ordered and regular MLG nanoclusters in *am*-II with reduced interlayer spacing and certain orientation (blue box area in Figure 2C), and the dense disordered regions become enlarged in *am*-II (red box area in Figure 2C). The corresponding FFT patterns of the marked HRTEM areas indicate that, unlike the obvious observed signal (inner halo) from disordered layered regions, the interlayer diffraction halo decreases until it almost disappears in the dense regions from *am*-I to *am*-II (Figures 2B to 2C), which can also been observed from their SAED patterns (Figures S5B and S5C). Thus, the *am*-I and *am*-II have the long-range disordered structure in common, but with different short-range orders. In the sample ⑤ synthesized at higher temperature, the dense disordered parts in *am*-II transformed to nanodiamond and compressed nanographite from the parts composed of more ordered MLG nanoclusters in *am*-II can be clearly seen in the composite (Figure 2D). The microstructure of this composite nanostructured material may be similar to that of natural impact diamonds and recently reported laboratory-synthesized diamond-related materials observed by HRTEM.[40–45] In these materials, the layered graphite-like domains observed are non-periodically inserted and coherently connected to the $sp^3$-bonded diamond domains. Several interface structure models of $sp^2$ (graphite-like) and $sp^3$ (diamond-like) units have been proposed to match the experimental HRTEM,[40,41,43–45] which deepens the understanding of the complex nanostructures in these diamond-related materials.

The $sp^2$ and $sp^3$ fractions of the *am*-I and *am*-II carbon materials were subsequently estimated by the carbon *K*-edge EELS spectra (Figures 3A). Compared with the low-loss EELS of raw $C_{60}$ centered at 26.0 eV, the low-loss EELS of the two amorphous carbon materials are right-shifted to 29.2 and 30.3 eV, respectively, indicating an increase in $sp^3$ bonding with the increase of synthesis temperature (Figure 3B). The core-loss EELS spectra display π* and σ* signatures corresponding to the transition of 1s to π* and 1s to σ* in the amorphous carbon materials,[46] respectively (Figure 3A). The π* component shows an obviously decrease from raw $C_{60}$ to the amorphous carbon materials, also indicating the increase in fraction of $sp^3$-bonded carbon atoms. The $sp^3$ contents of *am*-I and *am*-II are estimated at 52.8±2.9% and 64.9±3.3%, respectively. Furthermore, the previously calculation on Raman spectra of the amorphous carbon materials show that with the increase of tetrahedrally coordinated bonds in the disordered structure, the strong main peak of amorphous carbon gradually shifts from the position near the G-band of graphite to that near the D-band of diamond.[47] Notably, the C356 amorphous model of calculation containing 356 atoms of which 51.4% are tetrahedrally coordinated, has the main Raman peak position of 1374 cm$^{-1}$. Therefore, the rise of Raman peak near D-band of amorphous carbon materials we obtained should also related to the increase of internal $sp^3$ bonds, which is also consist with the EELS result. The densities of *am*-I and *am*-II are 2.68±0.05 and 3.09±0.08 g/cm$^3$, respectively, which are lower than that of diamond but higher than that (~2.0-2.5 g/cm$^3$) of the strong, hard, elastic and conductive compressed glassy carbon with $sp^3$ content up to 21%.[14]

**Structural Model of the Transformation**

We conducted the systematic study of the C$_{60}$ → amorphous carbon transformation that uncovered the mechanism governing material's evolution. It was demonstrated that increase of the synthesis pressure requires higher temperature for setting-in the diffusion process in the system to evolve towards certain metastable "equilibrium" state after the buckyballs crush. Combined with the XRD, Raman and HRTEM results, we propose the following phase transformation process at pressure of 15 GPa: with the synthesis temperature increase, the C$_{60}$ molecular spheres covalently bind to nearest neighbors to form 3D-C$_{60}$ polymer, and then, on further temperature increase to and reaching the molecules integrity threshold, the C$_{60}$ spheres in 3D-C$_{60}$ polymer gradually collapse into small fragments, that leads to the formation of amorphous carbon (*am*-I) phase. There are two kinds of disordered structures in the *am*-I, namely, disordered regions composed of short fairly straight fragments likely representing fused aromatic rings (nanographene "seeds") grown from the collapsed C$_{60}$ and denser, more abundant disordered areas displaying curled contrast characteristic of tetrahedrally bonded carbon atoms likely originating from the $sp^3$ linkages in the parent 3D-C$_{60}$ polymer. With the further increase of synthesis temperature, the short-curved fragments (the nano-graphene "seeds") merge into more ordered multi-layer graphene clusters of ~2 nm lateral (in-plane) size, and then, as temperature rises up, they develop into 5-10 nm "thick" MLG stacks with lateral dimension of 5-10 nm. At the same time, being more stable at high pressure the compact disordered regions rich in $sp^3$-bonded carbon gradually grow in volume, further densify due to continuous $sp^2$ → $sp^3$ conversion, and, finally, transform into nanodiamond clusters.

Based on our current understanding of high p,T synthesis using fullerene as precursor we propose an approach to design of novel amorphous carbon materials with controllable disorder. Specifically, in the considered class of systems the disorder originates in interplay between different types of short range order in the two-phase system. For example (at 15 GPa): "all-$sp^2$" structural units (short chains, fused aromatic rings, graphene nanoclusters) embedded into the denser disordered matrix with high concentration of the $sp^3$ cites (the second component/phase). The possibility of chemical tuning the $sp^2$/$sp^3$ ratio in the latter by altering the synthesis p,T offers a large variety of structural scenario/metastable phases with promising combination of physico-chemical properties to be explored. In this regard moving to pressure high enough to seize the nanographene clusters formation/growth from the aromatic rings created upon buckyballs crush on one hand and $sp^2$ → $sp^3$ carbon conversion/enriching the surrounding the dense disorderd matrix is undoubtedly the most intriguing and promising pathway in the future pursuit of new bulk, nearly pure

*sp*³/tetragonally coordinated amorphous carbon materials with enhanced physical properties.

**Mechanical Properties**

The hardness value of amorphous carbon materials produced at high p,T from $C_{60}$ reported earlier, still remains a controversial issue.[12,13,17,20,21] For example, it was argued that amorphous carbon synthesized at 13 GPa/1800 K is able to scratch diamond crystal surface thus possessing hardness exceeding that of diamond.[19] The hardness was evaluated according to Sclerometer test method, yielding the values as high as 170-300 GPa.[12] Later, the hardness measured by Vickers indentation method on the samples produced at similar conditions was only about 70 GPa.[13] To avoid such inconsistencies, we have measured the hardness of *am*-I and *am*-II by employing three independent hardness testing methods. As shown in Figure 4A and Figure S4, the asymptotic Vickers hardness ($H_V$) of *am*-I and *am*-II is 68.2±2.4 and 80.9±3.68 GPa, respectively whereas the Knoop hardness ($H_K$) reaches 43.4±2.7 and 56.7±1.0 GPa, respectively. The nano-indentation hardness ($H_N$) obtained by applying load in the range of 1.96-4.9 N is 64.4±1.7 and 76±1.3 GPa, for *am*-I and *am*-II, respectively, which is in good agreement with the Vickers hardness results (Figure 4A). Therefore, the *am*-I and *am*-II manifest hardness comparable to that of the (111) plane of single-crystal diamond.[48]

The Young's modulus (*E*) can be determined by the Oliver and Pharr method from the load/displacement curve during nano-indentation technique.[49] By assuming the Poisson's ratio of the materials is 0.2, the obtained *E* value of *am*-I and *am*-II is as high as 625±6.8 GPa and 874±24.2 GPa, respectively. The load-displacement curves exhibit also the high-elasticity response to local deformation of the two samples, and the elastic recovery after the load release is 80±1% and 69±1%, for *am*-I and *am*-II, respectively (Figure 4B). The high elasticity of amorphous carbon materials is due to the disorder and flexibility of nanometer-sized clusters comprised of several graphene layers (see Figure 2) and the existence of fingerprint-like curved microstructure in the former may be responsible for its higher elastic recovery.

The fracture toughness ($K_{Ic}$) of *am*-I and *am*-II was accurately determined from the cracks caused by Vickers indentation (Figures S6B and S6D). Remarkably, it was found that the two amorphous carbon materials have extremely high fracture toughness of 7.6±0.6 and 8.0±0.9 MPa·m$^{1/2}$, respectively (Figure 4C), which is very unusual for amorphous materials that normally exhibit low fracture toughness on the level of ~0.1-4.6 MPa·m$^{1/2}$.[50–54] Importantly, our results confirm the determined earlier with large uncertainly high fracture toughness value (8.7±1.5 MPa·m$^{1/2}$) of amorphous carbon produced from $C_{60}$ at lower pressure of 12.5 GPa and temperature of about 600 ℃. Comparison of the materials' fracture toughness with that of inorganic glasses and ta-C films is shown in Figure 4C. For example, the ordinary silicate glasses have $K_{Ic}$ of only ~1 MPa·m$^{1/2}$,[50] and the $K_{Ic}$ for a-$Al_2O_3$ and a-SiC films are 3.4 and 3.2 MPa·m$^{1/2}$,[55,56] respectively. Consequently, the amorphous carbon produced in this work have superior mechanical properties - hardness at the level of single-crystal and fracture toughness comparable to that of nanodiamond.[23]

**Electrical and Optical Properties**

There is only one report addressing the electrical properties of the amorphous carbon derived from fullerenes at p>10 GPa where very low bandgap (0.06-0.25 eV) was determined in the material recovered from pressures 12.5 GPa and 15 GPa.[22] Only two temperatures (827 ℃ and 1527 ℃) used in the synthesis at pressure of 15 GPa thus in the current study we extended the synthesis temperature range. The temperature dependence of the resistivity of the *am*-I and *am*-II samples, *ρ(*T*)*, exhibits behavior typical for semiconductor materials - resistivity decreases with temperature increase (Figure 5A). The room-temperature resistivity values are 19 and 35 ohm·cm for *am*-I and *am*-II, respectively, which is comparable to that of polycrystalline Ge semiconductor (~40 ohm·cm).[57] The approximate value of intrinsic bandgap of semiconductors can be calculated from the activation energy that is obtained from temperature dependence of conductivity σ(T) according to the equation[58]: $\sigma(T) = \sigma_0 \exp\left(-\frac{E_g}{2K_B T}\right)$, where $E_g$ is the bandgap, $\sigma_0$ is a constant, and $K_B$ is Boltzmann's constant. The bandgap values were also determined directly from the absorption edge of the FTIR-ATR spectra (Figure 5B) - 0.24 and 0.19 eV for *am*-I and *am*-II, respectively, that are very consistent with the results obtained from temperature-dependent resistivity.

The very-narrow bandgap observed in the synthesized amorphous materials is unusual for carbon systems. It is well known that diamond is an insulator with large bandgap of 5.5 eV whereas graphene has zero bandgap and C$_{60}$ is *sp$^2$*-type carbon that exhibits semiconducting behavior with a medium-size bandgap of ~1.5 eV. In addition, the 1D-, 2D-, 3D-polymerized C$_{60}$ are mostly electron conductive,[10] and the hard ta-C(:H) films have optical gaps of 0.4-3 eV (Figure 5C).[4,59,60] On the contrary, the amorphous carbon materials synthesized in bulk form in this work have not only excellent mechanical properties, (hardness comparable to-, and fracture toughness better than that of single-crystal diamond and large indentation elastic recovery) but also possess the bandgap range extending down to ~0.19 eV. This range corresponds to the infrared photon energy range, that covers not only the first atmospheric spectral window (~3-5 μm) in mid-infrared, but also extends to the second atmospheric spectral window (~8-14 μm) in far-infrared part of the spectrum thus making this kind of amorphous carbon materials a very promising material for mid-far infrared radiation detection and energy harvesting.

**Conclusions**

In summary, this work studied the transformation of fullerene $C_{60}$ at high pressure of 15 GPa and different synthesis temperatures, and clarified the microstructure and properties of resulting bulk amorphous carbon materials. To the best of our knowledge, this is the most comprehensive and systematic study of the structural transformation of fullerene to amorphous carbon that has been undertaken to date. For the first time all the transformation steps were followed through direct characterization of the structure on nanometer scale by HRTEM. That allowed us to build up a long-time missed generalized structural transformation model of $C_{60} \rightarrow$ amorphous carbon, ascertain the inconsistency in the results published earlier and propose future development of amorphous carbon materials through high p,T synthesis. We show that increase in the synthesis pressure, requires higher temperature for setting-in the diffusion process in the system to evolve towards certain metastable "equilibrium" state after the buckyballs crush. Such scenario of the system's microstructure evolution with synthesis temperature determines the difference in physical properties and performance of the resulting amorphous carbon materials. Particularly found that the amorphous carbon materials are actually narrow-band gap semiconductors with bandgaps only ~0.1-0.3 eV. This class of amorphous carbon materials not only have diamond-like hardness, but also have superior fracture toughness, which is comparable to that of nano-polycrystalline diamond.[23] This is very prominent in the amorphous system, which is about 8 times of the fracture toughness of silicate glass.[50] This kind of amorphous carbon materials with excellent mechanical properties and narrow bandgaps would have a potential to be used in many fields, for example, as superhard and high-toughness tools as well as unique/advanced photoelectric devices where diamond and graphene are not suitable.[61] Based on our current understanding of high p,T synthesis using fullerene as the precursor we proposed a different approach to design of novel amorphous carbon materials with controllable disorder compared to the packing disorder in layered $sp^2/sp^3$ polytypes.[43,44]

**EXPERIMENTAL METHOD**

**Sample Synthesis**

The samples were synthesized from fullerene $C_{60}$ powder (99.99%, Alfa Aesar) using the standard COMPRES 10/5 assembly for HPHT experiments in a large-volume multi-anvil press at Yanshan University. The raw $C_{60}$ powder was compacted and loaded into 2.0-mm inner diameter and 2.0-mm high h-BN capsules, and then assembled into a hole in the center of 10 mm spinel ($MgAl_2O_3$) octahedron with a Re heater and a $LaCrO_3$ thermal insulator. Pressure loading/unloading rates were set to 2 GPa/hour. Each sample was heated with a rate of 20 ℃/min to the peak temperature after the target pressure of 15 GPa was reached and thereafter quenched to ambient T after holding the peak temperature for 2 hours with subsequent pressure release. For each peak temperature (550°, 700°, 800°, 1000°, 1200 ℃), a new sample was prepared but the source fullerene material was taken from the same batch. The dimensions of the recovered specimens were up to ~1.5-1.7 mm in diameter and ~1.2-1.7 mm in height.

## X-ray Diffraction

The X-ray diffraction (XRD) spectra were taken with a Bruker D8 Discover diffractometer (Cu $K\alpha$ radiation) from the recovered bulk samples. *In-situ* angle dispersive X-ray diffraction (XRD) measurements were also performed at the 4W2 High-Pressure Station of Beijing Synchrotron Radiation Facility (BSRF) and BL15U1 Hard x-ray micro-focus beamline of Shanghai Synchrotron Radiation Facility (SSRF). The used X-ray wavelength is 0.6199 Å.

## Raman Spectroscopy

The Raman spectra were collected at ambient conditions using a Horiba Jobin–Yvon LabRAM HR-Evolution Raman microscope. The 325 nm and 532 nm lasers were used for excitation, the beam was focused on a spot of less than 2 μm in diameter on the sample surface. A special care was taken to avoid sample overheating during the spectra collection.

## HRTEM and EELS Measurements

Microstructures of the recovered samples were characterized by a condenser spherical aberration-corrected TEM (Themis Z, Thermo Fisher Scientific) with an accelerating voltage of 300 kV and TEM (Talos F200X) with an accelerating voltage of 200 kV. The specimens for high-resolution TEM (HRTEM) were prepared by a Ga focused ion beam (FIB, Scios Dual beam, Thermo Fisher Scientific) milling with an accelerating voltage of 30 kV from the bulk samples. To minimize the knock-out damage on specimens, the ion cleaning was executed at last by using a voltage of 5 kV and current of 16 pA for the electron-transparent slices with thickness of less than 100 nm. Electron energy-loss spectra (EELS) were collected in the TEM model from a randomly selected region of ~200 nm. The $sp^3$ ratio in carbon samples was estimated from the carbon K-edge EELS using the method,[14] and the raw $C_{60}$ from the same batch was used as an all-$sp^2$ (100%) the reference.

## Mechanical Performance Measurements

The hardness of the samples were investigated by means of three independent hardness measurements. The nanoindentation hardness ($H_N$) and Young's moduli ($E$) were derived from the load-displacement curves established by the three-sided pyramidal (Berkovich) diamond indenter (Nano Indenter G200). The applied loads range from 1.96 to 4.9 N, and the loading and dwelling times were both 15 s. The elastic recovery was calculated using the formula ($d_{max}$-$d_{min}$)/$d_{max}$, where $d_{max}$ and $d_{min}$ are the maximum displacement at maximum load and the residual after unloading, respectively. The Knoop ($H_K$) and Vickers ($H_V$) hardness measurements were carried out on the microhardness tester (KB 5 BVZ), and the adopted loading and dwelling times was 30 and 15 s, respectively. $H_K$ is determined from the equation $H_K = 14229P/d_1^2$, where $P$ (N) is the applied load and $d_1$ (μm) is the major diagonal length (long axis) of rhomboid-shaped Knoop indentation. $H_V$ is determined from the equation $H_V = 1854.4P/d_2^2$, where $d_2$ (μm) is the arithmetic mean of the two diagonals of Vickers indentation. The loads applied in $H_V$ and $H_K$ measurements were in the range of 0.98-6.86 N. Under each test method for each sample, at least five indentations were made at different loads. All the hardness values were determined from the SEM images of the indentations. The fracture toughness ($K_{Ic}$) was also estimated from Vickers indentation cracks produced at a load of 6.68 N. It was calculated from $K_{Ic} = 0.016(E/H_V)^{0.5}F/C^{1.5}$ for radial cracks formed on surfaces of bulk samples, where $C$ (in micrometers) is the average length of the radial cracks measured from the indent center, and $E$ is Young's modulus.

## Electrical Measurements

The electrical resistivity of the samples was measured in the temperature interval 100-360 K on a Physical Property Measurement System, Quantum Design, USA using the four-probe method. The platinum wires were adhered to the surface of the polished sample (~1.5 mm in diameter) with Leitsilber conductive silver cement (Ted Pella, silver content 45%).

## Infrared Absorption Measurements

The Attenuated Total Reflectance-Fourier transform infrared spectra (ATR-FTIR) were recorded in the spectral range of 600-4500 cm$^{-1}$ (E55+FRA106, Bruker) from the polished samples of ~1.5 mm in diameter and 0.5 mm in height at ambient conditions.

## DATA AVALIABILITY

**All data are reported in the paper or Supplemental Information.**

## SUPPLEMENTAL INFORMATION

Supplemental Information can be found online.

## ACKNOWLEDGEMENTS

This work was supported by the National Natural Science Foundation of China (Grants Nos. 52090020, 91963203, U20A20238, 51672238, 51722209), the National Key R&D Program of China (Grants No. 2018YFA0703400), the NSF for Distinguished Young Scholars of Hebei Province of China (Grant No. E2018203349) and Talent research project in Hebei Province (Grant No. 2020HBQZYC003).

## AUTHOR CONTRIBUTIONS

Z.Z., A.V.S. and Y.T. conceived the idea of this project; S.Z., B.L., K.L. and L.S. prepared the samples; S.Z., Y.W., Y. S., Y.Z. and K.L. measured the XRD and Raman spectra; S.Z., Y.G., P.Y.

and M.M. performed hardness measurements; S.Z., Y.W. and Y.S. scanned the indentations through the scanning electron microscope (SEM); S.Z. and B.L. measured the absorption spectra; K.L., S.Z., B.L. and Z.L. prepared the TEM samples using the focused ion beam (FIB) technology; K.L., W.H., S.Z., Y.W. and Z.L. conducted TEM and EELS characterization; Z.Z., S.Z., A.V.S., J.H., D.Y., B.X., Y.T., V.B., O.P.C., K.L., W.H., Y.W. and Y.G. analyzed the data; Z.Z. S.Z. and A.V.S drafted the manuscript with contributions from all authors. S.Z., Y.W., K.L. and B.L. contribute equally.

## DECLARATION OF INTERESTS

The authors declare no competing interests.

## REFERENCES

1. Schwierz, F. (2010). Graphene transistors. Nature Nanotech. *5*, 487–496.
2. Denis, P.A. (2010). Band gap opening of monolayer and bilayer graphene doped with aluminium, silicon, phosphorus, and sulfur. Chem. Phys. Lett. *492*, 251–257.
3. Zhang, Y., Tang, T.-T., Girit, C., Hao, Z., Martin, M.C., Zettl, A., Crommie, M.F., Shen, Y.R., and Wang, F. (2009). Direct observation of a widely tunable bandgap in bilayer graphene. Nature *459*, 820–823.
4. Robertson, J. (2002). Diamond-like amorphous carbon. Mater. Sci. Eng. R Rep *37*, 129–281.
5. Friedmann, T.A., Sullivan, J.P., Knapp, J.A., Tallant, D.R., Follstaedt, D.M., Medlin, D.L., and Mirkarimi, P.B. (1997). Thick stress-free amorphous-tetrahedral carbon films with hardness near that of diamond. Appl. Phys. Lett. *71*, 3820–3822.
6. Bewilogua, K., and Hofmann, D. (2014). History of diamond-like carbon films - From first experiments to worldwide applications. Surf. Coat. Technol. *242*, 214–225.
7. Bourdon, E.B.D., Duley, W.W., Jones, A.P., and Prince, R.H. (1991). Characterization of diamond-like films prepared by laser ablation of graphite. Surf. Caot. Technol. *47*, 509–516.
8. Dwivedi, N., Kumar, S., Malik, H.K., Govind, Rauthan, C.M.S., and Panwar, O.S. (2011). Correlation of $sp^3$ and $sp^2$ fraction of carbon with electrical, optical and nano-mechanical properties of argon-diluted diamond-like carbon films. Appl. Surf. Sci. *257*, 6804–6810.
9. Soldatov, A.V., Roth, G., Dzyabchenko, A., Johnels, D., Lebedkin, S., Meingast, C., Sundqvist, B., Haluska, M., and Kuzmany, H. (2001). Topochemical polymerization of $C_{70}$ controlled by monomer crystal packing. Science *293*, 680–683.
10. Álvarez-Murga, M., and Hodeau, J.L. (2015). Structural phase transitions of $C_{60}$ under high-pressure and high-temperature. Carbon *82*, 381–407.
11. Chernogorova, O., Potapova, I., Drozdova, E., Sirotinkin, V., Soldatov, A.V., Vasiliev, A., and Ekimov, E. (2014). Structure and physical properties of nanoclustered graphene synthesized from $C_{60}$ fullerene under high pressure and high temperature. Appl. Phys. Lett. *104*, 043110.
12. Blank, V.D., Buga, S.G., Dubitsky, G.A., Serebryanaya, N.R., PoPov, M.Y., and Sundqvist, B. (1998). High-pressure polymerized of $C_{60}$. Carbon *36*, 319–343.

13. Brazhkin, V.V., Solozhenko, V.L., Bugakov, V.I., Dub, S.N., Kurakevych, O.O., Kondrin, M.V., and Lyapin, A.G. (2007). Bulk nanostructured carbon phases prepared from $C_{60}$: approaching the 'ideal' hardness. J. Phys.: Condens. Matter *19*, 236209.

14. Hu, M., He, J., Zhao, Z., Strobel, T.A., Hu, W., Yu, D., Sun, H., Liu, L., Li, Z., Ma, M., et al. (2017). Compressed glassy carbon: An ultrastrong and elastic interpenetrating graphene network. Sci. Adv. *3*, e1603213.

15. Zeng, Z., Yang, L., Zeng, Q., Lou, H., Sheng, H., Wen, J., Miller, D.J., Meng, Y., Yang, W., Mao, W.L., et al. (2017). Synthesis of quenchable amorphous diamond. Nat. Commun. *8*, 322.

16. Hu, M., Zhang, S., Liu, B., Wu, Y., Luo, K., Li, Z., Ma, M., Yu, D., Liu, L., Gao, Y., et al. (2021). Heat-treated glassy carbon under pressure exhibiting superior hardness, strength and elasticity. J. Materiomics *7*, 177–184.

17. Talyzin, A.V., Langenhorst, F., Dubrovinskaia, N., Dub, S., and Dubrovinsky, L.S. (2005). Structural characterization of the hard fullerite phase obtained at 13 GPa and 830 K. Phys. Rev. B *71*, 115424.

18. Wood, R.A., Lewis, M.H., West, G., Bennington, S.M., Cain, M.G. and Kitamura, N. (2000). Transmission electron microscopy, electron diffraction and hardness studies of high-pressure and high-temperature treated $C_{60}$. J. Phys., Condens. Matter *12*, 10411–10421.

19. Chernogorova, O., Drozdova, E., Ovchinnikova, I., Soldatov, A.V. and Ekimov, E. (2012) Structure and properties of superelastic hard carbon phase created in fullerene-metal composites by high temperature-high pressure treatment, J. Appl. Phys. *111*, 112601.

20. Blank, V.D., Buga, S.G., Ivlev, A.N., and Mavrin, B.N. (1995). Ultrahard and superhard carbon phases produced from $C_{60}$ by heating at high pressure:, structural and Raman studies. Phys. Lett. A *205*, 208–216.

21. Brazhkin, V.V., and Lyapin, A.G. (2012). Hard and superhard carbon phases synthesized from fullerites under pressure. J. Superhard Mater. *34*, 400–423.

22. Buga, S.G., Blank, V.D., Serebryanaya, N.R., Dzwilewski, A., Makarova, T. and Sundqvist, B. (2005). Electrical properties of 3D-polymeric crystalline and disordered $C_{60}$ and $C_{70}$ fullerites. Diam. Relat. Mater. *14*, 896–901.

23. Mohr, M., Caron, A., Herbeck-Engel, P., Bennewitz, R., Gluche, P., Brühne, K. and Fecht, H.-J. (2014). Young's modulus, fracture strength, and Poisson's ratio of nanocrystalline diamond films. J. Appl. Phys. *116*, 124308.

24. Blank, V.D., Dubitsky, G.A., Serebryanaya, N.R., Mavrin, B.N., Denisov, V.N., Buga, S.G., and Chernozatonskii, L.A. (2003). Structure and properties of $C_{60}$ and $C_{70}$ phases produced under 15 GPa pressure and high temperature. Phys. B:, Condens. Matter *339*, 39–44.

25. Shibazaki, Y., Kono, Y., and Shen, G. (2019). Compressed glassy carbon maintaining graphite-like structure with linkage formation between graphene layers. Sci. Rep. *9*, 7531.

26. Álvarez-Murga, M., Bleuet, P., Garbarino, G., Salamat, A., Mezouar, M., and Hodeau, J.L. (2012). "Compressed graphite" formed during $C_{60}$ to diamond transformation as revealed by scattering computed tomography. Phys. Rev. Lett. *109*, 025502.

27. Bethune, D.S., Meijer, G., Tang, W.C., Rosen, H.J., Golden, W.G., Seki, H., Brown, C.A., and de Vries, M.S. (1991). Vibrational Raman and infrared spectra of chromatographically separated $C_{60}$ and $C_{70}$ fullerene clusters. Chem. Phys. Lett. *179*, 7.

28. Ferrari, A.C., and Robertson, J. (2000). Interpretation of Raman spectra of disordered and amorphous carbon. Phys. Rev. B *61*, 14095–14107.
29. Ferrari, A.C., and Robertson, J. (2001). Resonant Raman spectroscopy of disordered, amorphous, and diamondlike carbon. Phys. Rev. B *64*, 075414.
30. Stein, I.Y., Constable, A.J., Morales-Medina, N., Sackier, C.V., Devoe, M.E., Vincent, H.M. and Wardle, B.L. (2017). Structure-mechanical property relations of non-graphitizing pyrolytic carbon synthesized at low temperatures. Carbon *117*, 411–420.
31. Mallet-Ladeira, P., Puech, P., Toulouse, C., Cazayous, M., Ratel-Ramond, N., Weisbecker, P., Vignoles, G.L. and Monthioux, M. (2014). A Raman study to obtain crystallite size of carbon materials, A better alternative to the Tuinstra–Koenig law. Carbon *80*, 629–639.
32. Smith, M.W. and Tu, K.N. (2016). Structural analysis of char by Raman spectroscopy: Improving band assignments through computational calculations from first principles. J. Appl. Phys. *15*, 2370-2370
33. Wang, Q.; Wang, C.; Wang, Z.; Zhang, J. and He, D. (2007). Fullerene nanostructure-induced excellent mechanical properties in hydrogenated amorphous carbon. Appl. Phys. Lett. *91*, 141902.
34. Gupta, A.K., Tang, Y., Crespi, V.H. and Eklund, P.C. (2010). A non-dispersive Raman D-band activated by well-ordered interlayer interactions in rotationally stacked bi-layer Graphene. Phys. Rev. B *82*, 241406(R)
35. Milani, P., Ferretti, M., Piseri, P., Bottani, C.E.; Ferrari, A., Li Bassi, A., Guizzetti, G. and Patrini, M. (1997). Synthesis and characterization of cluster-assembled carbon thin films. J. Appl. Phys. *82*, 5793–5798.
36. Ferrari, A.C. and Robertson, J. (2001). Origin of the 1150−cm$^{-1}$ Raman mode in nanocrystalline diamond, Phys. Rev. B *63*, 121405(R)
37. Hasdeo, E.H., Nugraha, A.R.T., Dresselhaus, M.S.and Saito, R. (2014). Breit-Wigner-Fano line shapes in Raman spectra of graphene. Phys. Rev. B, *90*, 245140.
38. Ribeiro-Soares, J., Oliveros, M.E., Garin, C., David, M.V., Martins, L.G.P., Almeida, C.A., Martins-Ferreira, E.H., Takai, K., Enoki, T., et al. (2015). Structural analysis of polycrystalline graphene systems by Raman spectroscopy. Carbon *95*, 646–652.
39. Pimenta, M.A., Dresselhaus, G., Dresselhaus, M.S., Cançado, L.G., Jorio, A. and Saito, R. (2007). Studying disorder in graphite-based systems by Raman spectroscopy.Phys. Chem. Chem. Phys. *9*, 1276–1290.
40. Garvie, L.A.J., Nemeth, P., and Buseck, P.R. (2014). Transformation of graphite to diamond via a topotactic mechanism. Am. Min. *99*, 531–538.
41. Zhang, S., Zhang, Q., Liu, Z., Legut, D., Germann, T.C., Veprek, S., Zhang, H., and Zhang, R. (2020). Ultrastrong π-bonded interface as ductile plastic flow channel in nanostructured diamond. ACS Appl. Mater. Interfaces *12*, 4135–4142.
42. Xie, Y.-P., Zhang, X.-J., and Liu, Z.-P. (2017). Graphite to diamond:, origin for kinetics selectivity. J. Am. Chem. Soc. *139*, 2545–2548.
43. Németh, P., McColl, K., Garvie, L.A.J., Salzmann, C.G., Murri, M., and McMillan, P.F. (2020). Complex nanostructures in diamond. Nat. Mater. *19*, 1126–1131.

44. Németh, P., McColl, K., Smith, R.L., Murri, M., Garvie, L.A.J., Alvaro, M., Pécz, B., Jones, A.P., Corà, F., Salzmann, C.G., et al. (2020). Diamond-graphene composite nanostructures. Nano Lett. *20*, 3611–3619.

45. Zhu, S., Yan, X., Liu, J., Oganov, A.R., and Zhu, Q. (2020). A revisited mechanism of the graphite-to-diamond transition at high temperature. Matter *3*, 864–878.

46. Fallon, P.J., Veerasamy, V.S., Davis, C.A., Robertson, J., Amaratunga, G.A.J., Milne, W.I., and Koskinen, J. (1993). Properties of filtered-ion-beam-deposited diamondlike carbon as a function of ion energy. Phys. Rev. B *48*, 4777–4782.

47. Beeman, D., Silverman, J., Lynds, R., and Anderson, M.R. (1984). Modeling studies of amorphous carbon. Phys. Rev. B *30*, 870–875.

48. Brookes, C.A., and Brookes, E.J. (1991). Diamond in perspective: A review of mechanical properties of natural diamond. Diam. Relat. Mater. *1*, 13–17.

49. Oliver, W.C., and Pharr, G.M. (1992). An improved technique for determining hardness and elastic modulus using load and displacement sensing indentation experiments. J. Mater. Res. *7*, 1564–1583.

50. Hand, R.J., and Tadjiev, D.R. (2010). Mechanical properties of silicate glasses as a function of composition. J. Non-Cryst. Solids *356*, 2417–2423.

51. Baik, D.S., No, K.S., Chun, J.S., Yoon, Y.J., and Cho, H.Y. (1995). A comparative evaluation method of machinability for mica-based glass-ceramics. J. Mater. Sci. *30*, 1801–1806.

52. Bauer, A., Christ, M., Zimmermann, A., and Aldinger, F. (2001). Fracture toughness of amorphous precursor‐derived ceramics in the silicon‐carbon‐nitrogen system. J. Am. Chem. Soc. *84*, 5.

53. Guin, J.-P., Rouxel, T., Sanglebœuf, J.-C., Melscoët, I., and Lucas, J. (2002). Hardness, toughness, and scratchability of germanium-selenium chalcogenide glasses. J. Am. Chem. Soc. *85*, 1545–1552.

54. Rouxel, T., and Yoshida, S. (2017). The fracture toughness of inorganic glasses. J. Am. Ceram. Soc. *100*, 4374–4396.

55. Xia, Z., Riester, L., Sheldon, B.W., Curtin, W.A., Liang, J., Yin, A., and Xu, J.M. Mechanical properties of highly ordered nanoporous anodic alumina membranes. Rev. Adv. Mater. Sci. *6*, 131–139.

56. Zorman, C.A., and Parro, R.J. (2008). Micro‐ and nanomechanical structures for silicon carbide MEMS and NEMS. Phys. Stat. Sol. (B) *245*, 1404–1424.

57. Steele, M.C., and Rosi, F.D. (1958). Thermal conductivity and thermoelectric power of germanium‐silicon alloys. J. Appl. Phys. *29*, 1517–1520.

58. Zhao, Z., Zhang, H., Kim, D.Y., Hu, W., Bullock, E.S., and Strobel, T.A. (2017). Properties of the exotic metastable ST12 germanium allotrope. Nat. Commun. *8*, 13909.

59. Coşkun, Ö.D., and Zerrin, T. (2015). Optical, structural and bonding properties of diamond-like amorphous carbon films deposited by DC magnetron sputtering. Diam. Relat. Mater. *56*, 29–35.

60. Savvides, N. (1986). Optical constants and associated functions of metastable diamondlike amorphous carbon films in the energy range 0.5-7.3 eV. J. Appl. Phys. *59*, 4133–4145.

61. Chu, J., and Sher, A. (2010). Device physics of narrow gap semiconductors (New York:, Springer).

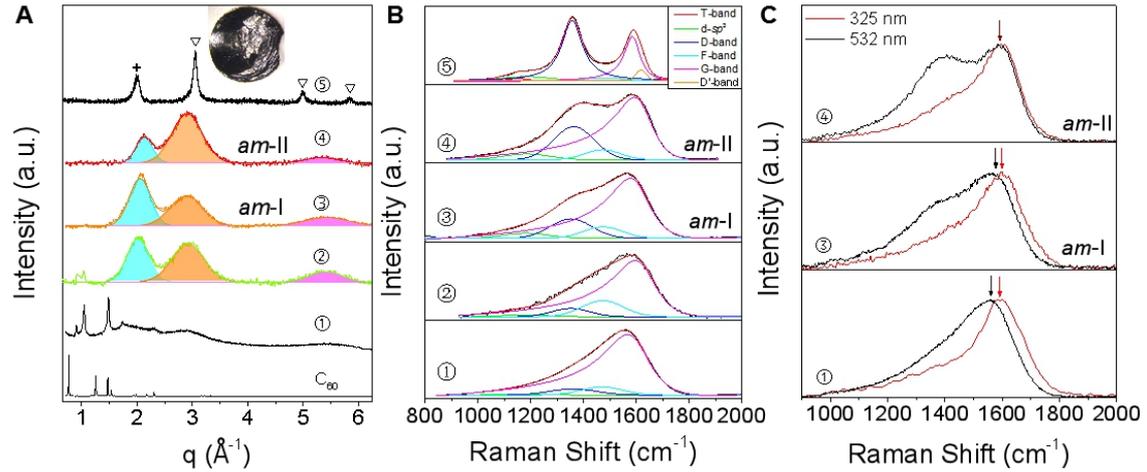

**Figure 1. XRD Patterns and Raman Spectra of Resulting Carbon Materials Measured at Ambient Conditions.** The numbers ①-⑤ represent the resulting carbon materials from compressing $C_{60}$ at 15 GPa and various temperatures of 550°, 700°, 800°, 1000°, and 1200 ℃, respectively. All the recovered samples are black blocks, and the inset in (A) is a photo of *am*-I with diameter of ~1.6 mm.

(A) The XRD patterns indicates that $C_{60}$ is first transformed to 3D-$C_{60}$ at high pressure and high temperature, and then begins to amorphization, and finally changes to compressed graphite-diamond composite. The amorphous carbon materials, depicted as *am*-I and *am*-II, have three broad diffraction peaks with similar positions at q = ~2.1, ~3.0 and ~5.4 Å$^{-1}$ but different intensities. The star and triangles are from the reflections of compressed graphite and diamond, respectively

(B) Raman spectra of samples ①-⑤ collected using 532 nm laser excitation. Peak fitting of the Raman spectra according to Ferrari & Robertson model[28,29,36] with position specific to our systems. The peak assignment are listed in Table S1 and peak decomposition of the spectra are summarized in Table 1.

(C) Comparison of Raman spectra excited by 325 and 532 nm laser. The comparison shows a dispersion of G-band in samples ①, ③ and ④, which means the presence of different types of disordered components in the microstructure. All Raman data are background free.

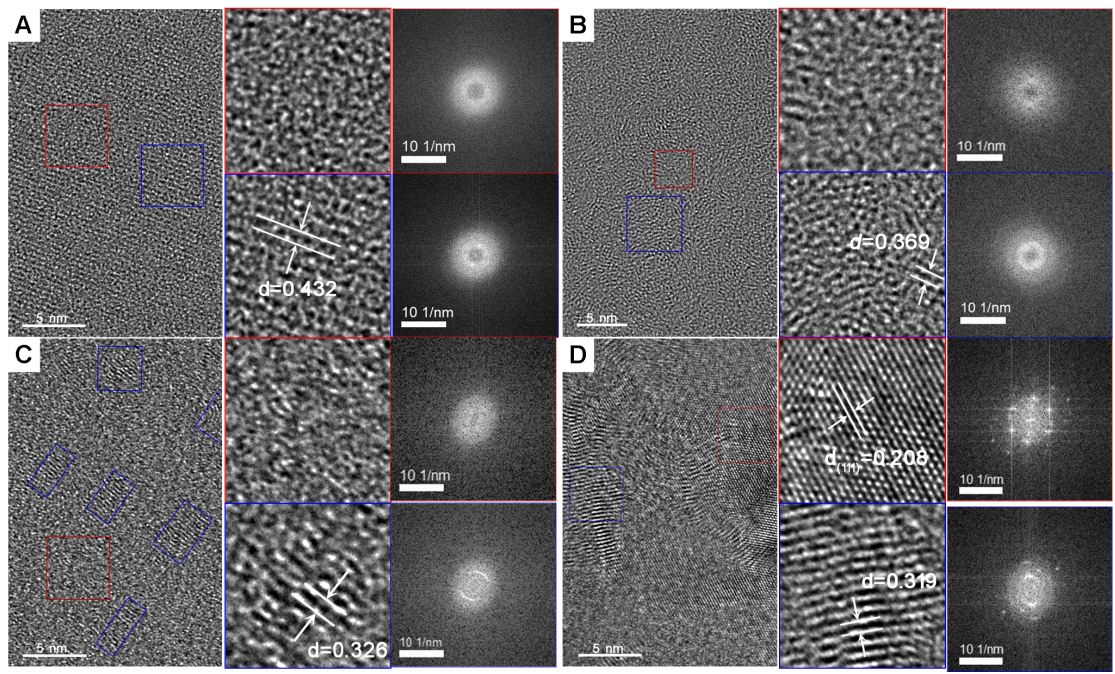

**Figure 2. HRTEM Images of Samples ①, ③, ④, and ⑤ Corresponding to 3D-C$_{60}$ with Partial Amorphization, *am*-I, *am*-II, and Compressed Nanographite-nanodiamond Composite.** The figures on the right in each panel include the zoomed-in view of the square-marked areas in the HRTEM images.

(A) The crystalline periodicity of the residual 3D-C$_{60}$ (blue box area) is interrupted by the amorphous components (red box area). The corresponding FFT patterns show the reveal crystalline spots embedded in the amorphous halo.

(B) The *am*-I shows the completely disordered characteristics composed of fingerprint-like small curved nanographene fragments (blue box area) and more compact disordered structure (red box area). The corresponding FFT patterns in two regions both exhibit diffuse rings.

(C) The compact disordered components in *am*-II are enlarged and almost connected together (red box area), and the residual multilayer graphene clusters from the merge of the short curved nanographene fragments become ordered and oriented with reduced interlayer spacing (blue box area). The FFT pattern in blue box area also shows the orientation of multilayer graphene component in *am*-II, and in comparison, the interlayer diffraction signal almost disappeared in the more compact regions.

(D) The nanodiamond (red box area) and compressed graphite (blue box area) are clearly seen in the composite.

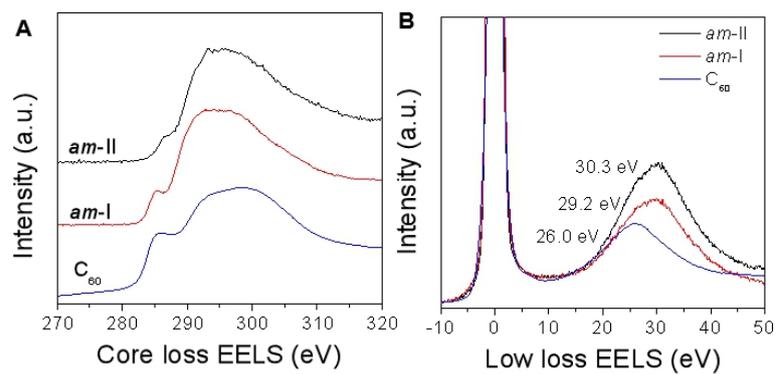

**Figure 3. EELS Spectra of Pristine $C_{60}$ and Resulting *am*-I and *am*-II Samples.**

(A) The core loss EELS showing the decreased of $sp^2$ contribution in the *am*-I and *am*-II relative to the pristine $C_{60}$.

(B) The right shift of the plasmon peak in low loss EELS spectra of *am*-I and *am*-II in comparison to pristine $C_{60}$ indicates an increase of $sp^3$ bonded carbons in structure.

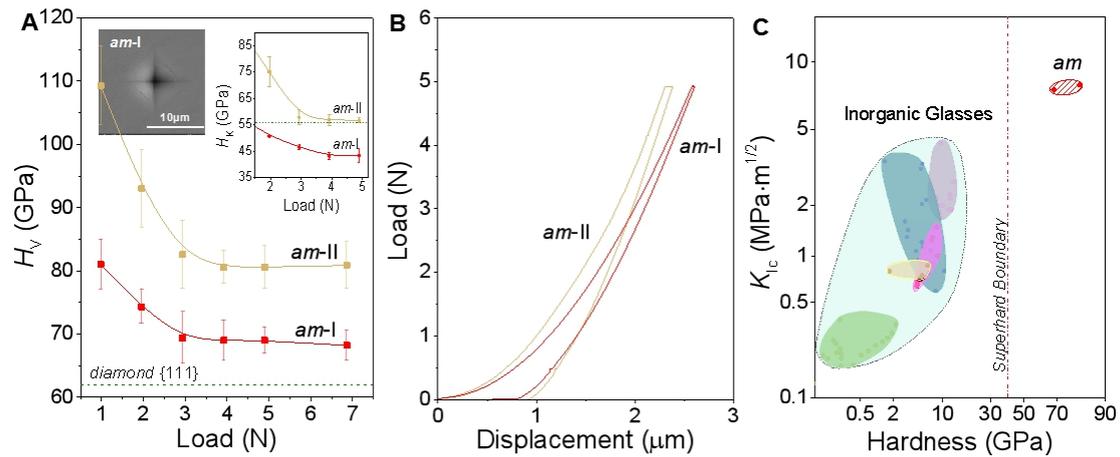

**Figure 4. Mechanical Properties of *am*-I and *am*-II.**

(A) Vickers hardness ($H_V$) of *am*-I and *am*-II as a function of applied loads. At loads exceeding 2.94 N, $H_V$ approaches the asymptotic value. Left panel inset: SEM image of the residual indentation made by the Vickers pyramid probe on the *am*-I surface after unloading from 4.9 N. Right panel inset: Knoop hardness ($H_K$) of *am*-I and *am*-II as a function of applied loads. Error bars indicate s.d. (n=5). The dashed lines indicate $H_V$ and $H_K$ of (111) plane of natural diamond crystal.[48]

(B) The load-displacement curves of *am*-I and *am*-II recorded during nanoindentation testing, exhibiting high elastic recovery.

(C) Comparison of hardness and fracture toughness between *am*s (current work) and ordinary inorganic glasses including silicate glasses (dark cyan),[50] $B_2O_3$-based glasses (magenta),[51] silicon oxynitride glasses (yellow),[52] chalcogenide glasses (green),[53] and $SiO_2$-based glasses (purple).[54]

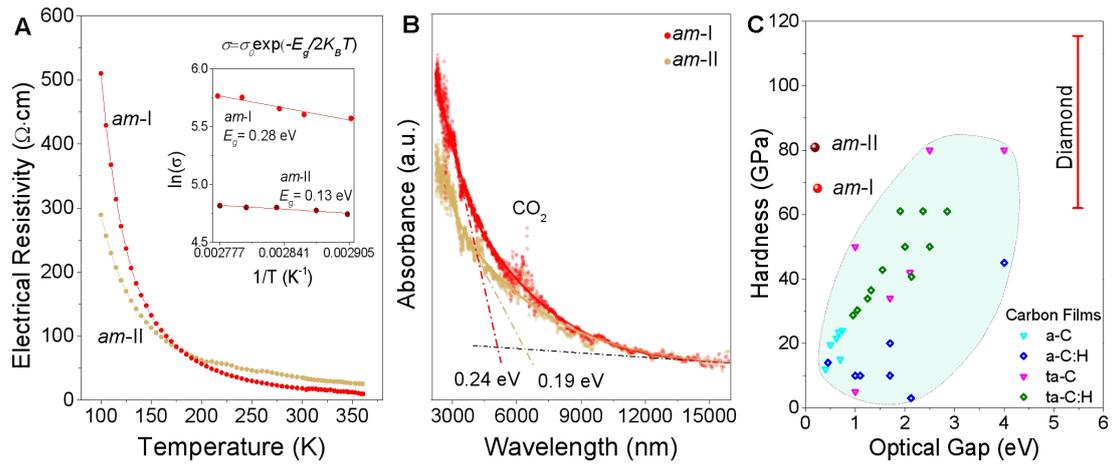

**Figure 5. Electrical and Optical Properties of *am*-I and *am*-II.**

(A) Temperature dependence of the electrical resistivity shows the semiconductor behavior of *am*-I and *am*-II. Inset: the activation energies obtained from the relationship of high-temperature conductivity. The estimated bandgaps of *am*-I and *am*-II are 0.28 and 0.13 eV, respectively.

(B) ATR-FTIR spectra of *am*-I and *am*-II. The absorption edges of *am*-I and *am*-II are located at ~5100 and ~6500 nm, respectively, corresponding to the bandgaps of 0.24 and 0.19 eV. The sharp peaks around ~6400 nm is caused by $CO_2$ in air.

(C) Hardness and optical gap of *am*-I and *am*-II in comparison with (non-)hydrogenated amorphous carbon films[4,59,60] and single-crystal diamond.[48]

Table 1. Summary of the Raman spectra analysis (the key parameters derived from the spectra): G-band peak position, width (FWHM) and shape (asymmetry factor $Q_{BWF}$), $1/Q_{BWF}$ parameter, D/G(BWF) peak area ratio and coherent scatterers' size ($L_a$).

| Temperature (℃) | G-band position (cm$^{-1}$) | G-band width (FWHM) | D/G (peak area), % | $Q_{BWF}$ | $1/Q_{BWF}$ | Cluster size $L_a$(nm) |
|---|---|---|---|---|---|---|
| **550** | 1567 | 230 | 0.07 | -3.57 | -0.28 | <2 |
| **700** | 1577 | 218 | 0.06 | -3.70 | -0.27 | <2 |
| **800** | 1578 | 210 | 0.10 | -3.45 | -0.29 | <2 |
| **1000** | 1593 | 196 | 0.38 | -3.23 | -0.31 | <2 |
| **1200** | 1586 | 74 | 1.89 | -12.5 | -0.08 | 10±2 |

# Supplemental Information

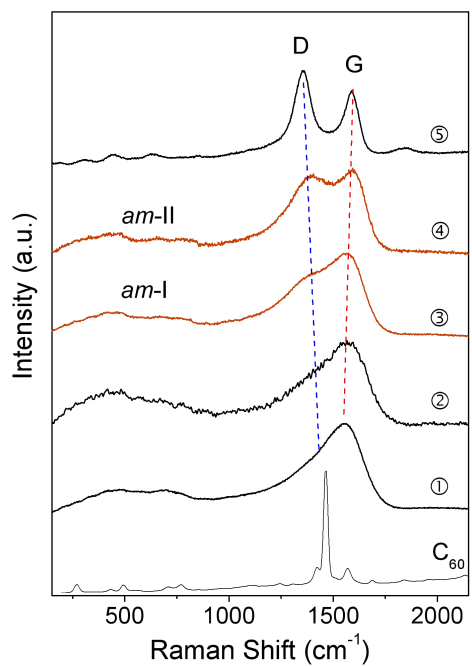

**Figure S1. Raman spectra of resulting carbon materials excited by 532 nm laser.** There is a wide asymmetric band with strongest position ~1560 cm$^{-1}$ in samples ① and ②. With the synthesis temperature increase, another broad band at the left shoulder of the asymmetric band can be visible in samples ③ and ④ (i.e. *am*-I and *am*-II), and the additional band intensity increases with synthesis temperature. For the compressed graphite-diamond composite sample ⑤, the D- and G-bands can be distinguished. The dashed lines give an indication of the peak shifts.

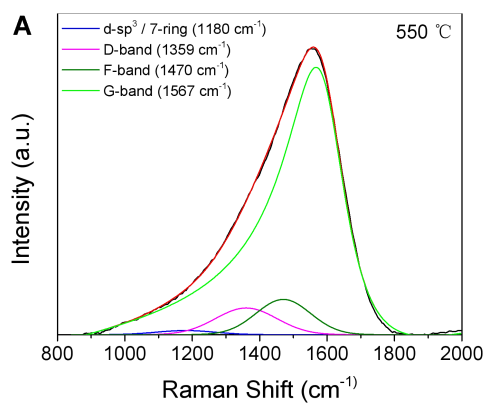

| λ 532 nm | Band | Raman Shift (cm$^{-1}$) | Intensity (Peak Area) | Intensity (Peak Amplitude) | FWHM (cm$^{-1}$) |
|---|---|---|---|---|---|
| 550 °C | d-sp$^3$/7-Ring | 1180 | 0.97 | 0.03 | 205 |
| | D | 1360 | 6.58 | 0.10 | 213 |
| | F | 1470 | 7.73 | 0.20 | 190 |
| | G | 1567 | 84.72 | 1.87 | 230 |

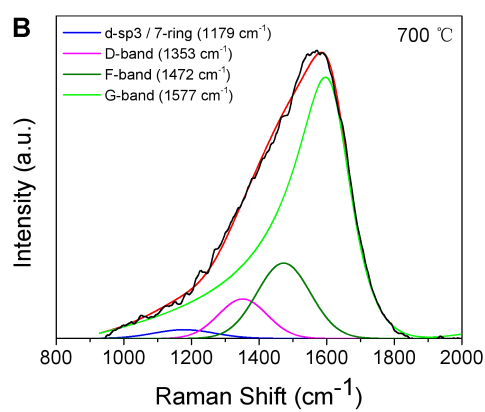

| λ 532 nm | Band | Raman Shift (cm$^{-1}$) | Intensity (Peak Area) | Intensity (Peak Amplitude) | FWHM (cm$^{-1}$) |
|---|---|---|---|---|---|
| 700 °C | d-sp$^3$/7-Ring | 1179 | 0.96 | 0.04 | 144 |
| | D | 1353 | 5.58 | 0.20 | 149 |
| | F | 1472 | 7.45 | 0.30 | 124 |
| | G | 1577 | 86.01 | 1.96 | 218 |

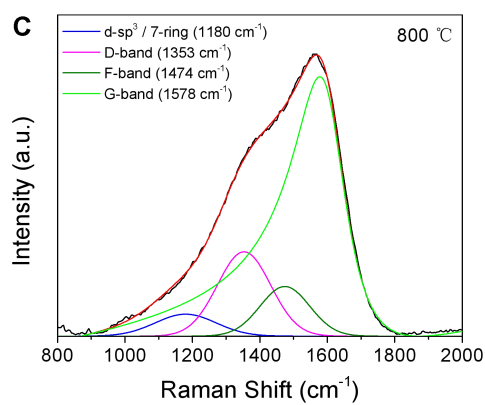

| λ 532 nm | Band | Raman Shift (cm$^{-1}$) | Intensity (Peak Area) | Intensity (Peak Amplitude) | FWHM (cm$^{-1}$) |
|---|---|---|---|---|---|
| 800 °C | d-sp$^3$/7-Ring | 1180 | 2.51 | 0.15 | 212 |
| | D | 1353 | 8.44 | 0.60 | 186 |
| | F | 1474 | 4.62 | 0.30 | 173 |
| | G | 1578 | 84.43 | 1.81 | 210 |

**Figure S2.** Peak fitting of the Raman spectra collected using 532 nm laser excitation from the amorphous carbon materials synthesized at 15 GPa and temperature of 550°, 700°, and 800 ℃, panels A, B, C, respectively, with the correspondent tables summarizing peak decomposition of the spectra.

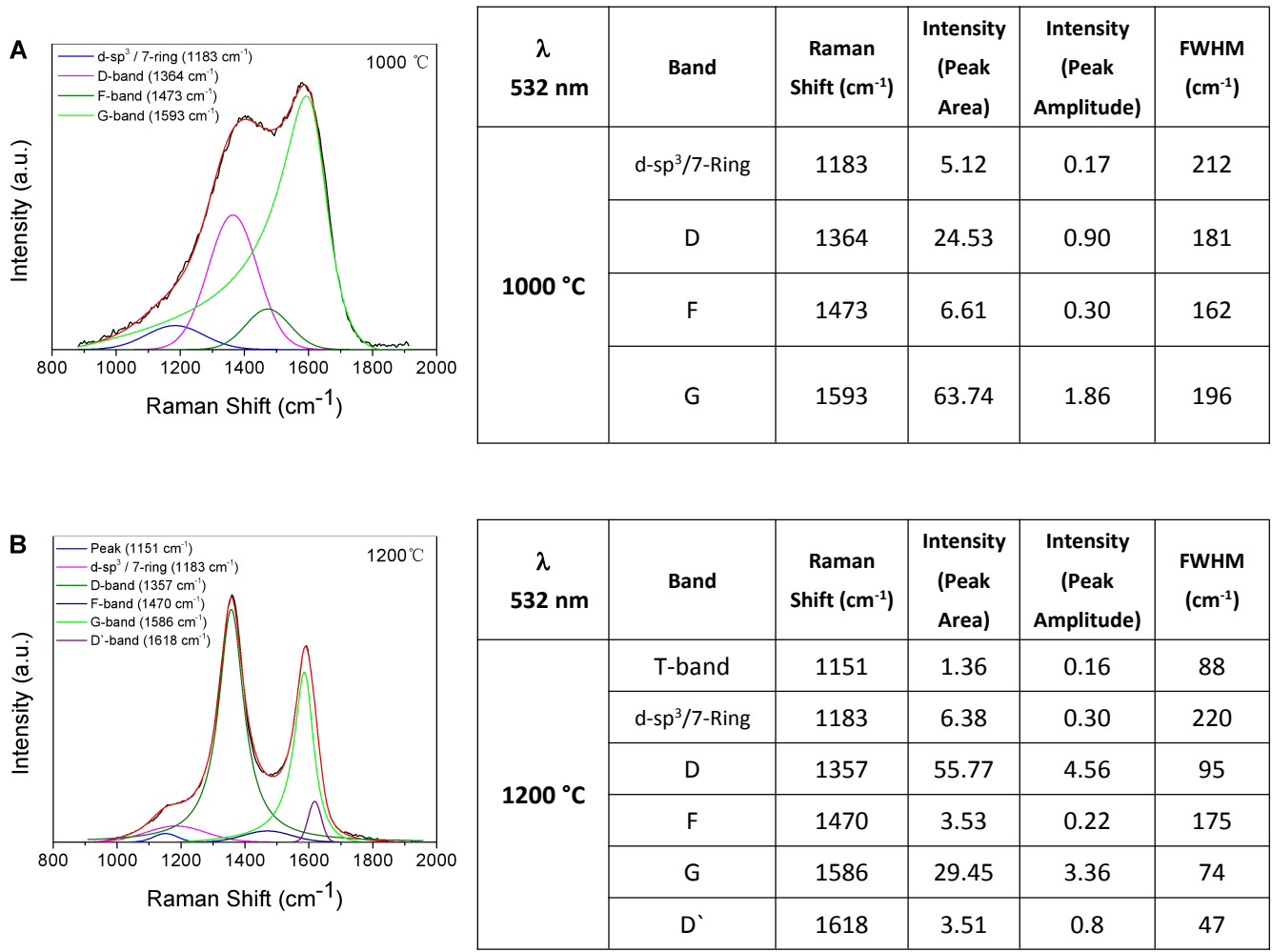

| λ 532 nm | Band | Raman Shift (cm$^{-1}$) | Intensity (Peak Area) | Intensity (Peak Amplitude) | FWHM (cm$^{-1}$) |
|---|---|---|---|---|---|
| 1000 °C | d-sp$^3$/7-Ring | 1183 | 5.12 | 0.17 | 212 |
| | D | 1364 | 24.53 | 0.90 | 181 |
| | F | 1473 | 6.61 | 0.30 | 162 |
| | G | 1593 | 63.74 | 1.86 | 196 |

| λ 532 nm | Band | Raman Shift (cm$^{-1}$) | Intensity (Peak Area) | Intensity (Peak Amplitude) | FWHM (cm$^{-1}$) |
|---|---|---|---|---|---|
| 1200 °C | T-band | 1151 | 1.36 | 0.16 | 88 |
| | d-sp$^3$/7-Ring | 1183 | 6.38 | 0.30 | 220 |
| | D | 1357 | 55.77 | 4.56 | 95 |
| | F | 1470 | 3.53 | 0.22 | 175 |
| | G | 1586 | 29.45 | 3.36 | 74 |
| | D` | 1618 | 3.51 | 0.8 | 47 |

**Figure S3.** Peak fitting of the Raman spectra collected using 532 nm laser excitation from the amorphous carbon materials synthesized at 15 GPa and temperature of 1000°, and 1200 ℃, panels A and B, respectively, with the correspondent tables summarizing peak decomposition of the spectra.

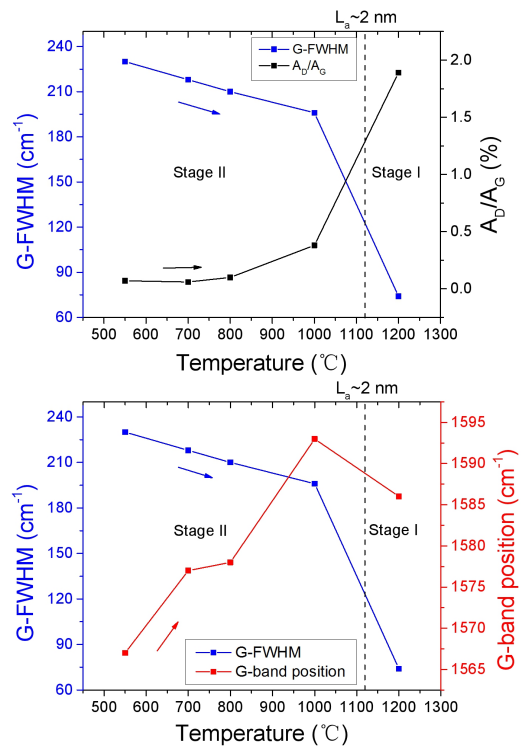

**Figure S4. BWF/G band peak width (G-FWHM) and the D/G peak area ratio ($A_D/A_G$) dependence on the synthesis temperature of samples.** Nanographene clusters' size ($L_a$) was estimated from the G-band FWHM[1] and D/G peak area ratio.[2]

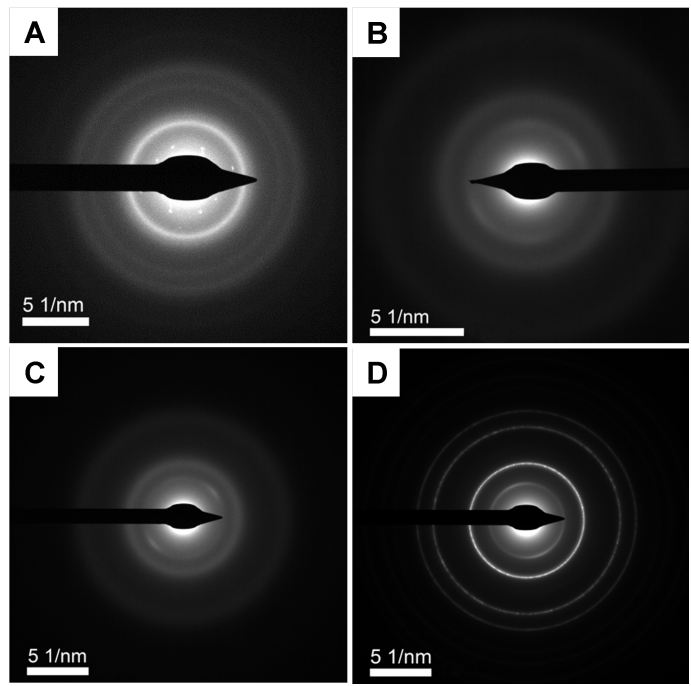

**Figure S5. SAED patterns of samples ①, ③, ④, and ⑤ corresponding Figure 2A-D.** The diffraction patterns were collected from linear dimensions of ~200 nm.

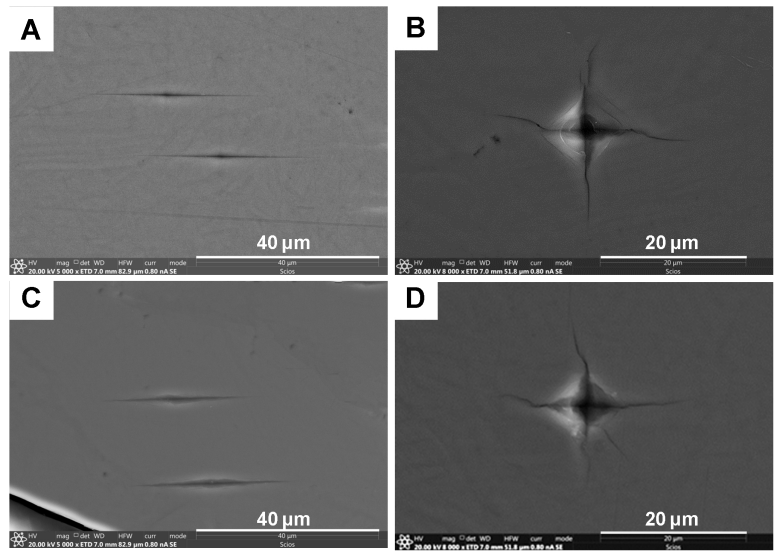

**Figure S6. Knoop and Vickers indentations after unloading.** (A) and (C) SEM images of Knoop indentations of *am*-I and *am*-II materials formed by a load of 4.9 N, respectively. (B) and (D) SEM images of Vickers indentations with cracks on *am*-I and *am*-II surface after unloading from 6.86 N, respectively. All images were recorded after the load release.

**Table S1. Peak assignment used in decomposition of the Raman spectra in Figure 1B.**

| Vibrational mode (band) | Raman Shift (cm$^{-1}$) | Reference |
|:---:|:---:|:---:|
| T (*sp$^3$* carbon) | 900-1300 | 3 |
| d-*sp$^3$* | 1180 | 4,5 |
| 7-ring | 1150-1200 | 6–8 |
| D (hexagonal rings) | 1355-1365 (532 nm) | 3,6,9–12 |
| F (5-ring) | 1450-1470 | 4–6,11,12 |
| G (*sp$^2$* carbon) | 1570-1600* | 3,9–11 |
| D' (graphene) | 1620 | 3,6,9–11 |

*Peak assigned according to Ferrari & Robertson model with position specific to our systems.

**Reference**

1. Ribeiro-Soares, J. *et al.* Structural analysis of polycrystalline graphene systems by Raman spectroscopy. *Carbon* **95**, 646–652 (2015).

2. Pimenta, M. A. *et al.* Studying disorder in graphite-based systems by Raman spectroscopy. *Phys. Chem. Chem. Phys.* **9**, 1276–1290 (2007).

3. Ferrari, A. C. & Robertson, J. Resonant Raman spectroscopy of disordered, amorphous, and diamondlike carbon. *Phys. Rev. B* **64**, (2001).

4. Milani, P. *et al.* Synthesis and characterization of cluster-assembled carbon thin films. *J. Appl. Phys.* **82**, 5793–5798 (1997).

5. Ferrari, A. C. & J. Robertson. Origin of the 1150−cm$^{-1}$ Raman mode in nanocrystalline diamond. *Phys. Rev. B* **63**, 121405(R) (2001).

6. Chernogorova, O. *et al.* Structure and physical properties of nanoclustered graphene synthesized from $C_{60}$ fullerene under high pressure and high temperature. *Appl. Phys. Lett.* **104**, 043110 (2014).

7. Smith, M. W. Structural analysis of char by Raman spectroscopy: Improving band assignments through computational calculations from first principles. *Carbon* **100**, 378–392 (2016).

8. Wang, Q., Wang, C., Wang, Z., Zhang, J. & He, D. Fullerene nanostructure-induced excellent mechanical properties in hydrogenated amorphous carbon. *Appl. Phys. Lett.* **91**, 141902 (2007).

9. Ferrari, A. C. & Robertson, J. Interpretation of Raman spectra of disordered and amorphous carbon. *Phys. Rev. B* **61**, 14095–14107 (2000).

10. Stein, I. Y. *et al.* Structure-mechanical property relations of non-graphitizing pyrolytic carbon synthesized at low temperatures. *Carbon* **117**, 411–420 (2017).

11. Mallet-Ladeira, P. *et al.* A Raman study to obtain crystallite size of carbon materials: A better alternative to the Tuinstra–Koenig law. *Carbon* **80**, 629–639 (2014).

12. Gupta, A. K., Tang, Y., Crespi, V. H. & Eklund, P. C. A non-dispersive Raman D-band activated by well-ordered interlayer interactions in rotationally stacked bi-layer Graphene. *Phys. Rev. B* **82**, 241406(R) (2010).